\documentclass[aps,prl,twocolumn,superscriptaddress,longbibliography,hidelinks]{revtex4-2}

\usepackage{color}
\usepackage{graphicx}
\usepackage{siunitx}
\usepackage{amsmath}
\usepackage{hyperref}

\begin{document}

\title{Rheologically tuned modes of collective transport in active viscoelastic films}

\author{Henning Reinken}
\email{henning.reinken@ovgu.de}
\affiliation{Institut f\"ur Physik, Otto-von-Guericke-Universität Magdeburg, Universitätsplatz 2, 39106 Magdeburg, Germany} 

\author{Andreas M. Menzel}
\email{a.menzel@ovgu.de}
\affiliation{Institut f\"ur Physik, Otto-von-Guericke-Universität Magdeburg, Universitätsplatz 2, 39106 Magdeburg, Germany}

\date{\today}

\begin{abstract}
While many living biological media combine both viscous and elastic properties, most theoretical studies employ either purely fluid- or solid-like descriptions.
We here use a unified framework for active films on substrates capable of describing a broad range of viscoelastic behavior to explore the interplay between activity and rheology.
The core of the study is a comprehensive state diagram showing a rich world of spatiotemporal dynamic states. %
Our results demonstrate the potential of tunable rheology to realize modes of controlled active transport on the microscale. 
\end{abstract}

\maketitle

Many living biological materials simultaneously exhibit properties of both viscous fluids and elastic solids~\cite{li2021microswimming} or may even change their behavior dynamically.
For example, during biofilm formation, bacteria excrete extracellular polymeric substances~\cite{hall2004bacterial,worlitzer2022biophysical}, which results in complex rheological behavior~\cite{jana2020nonlinear}.
Depending on the stage of biofilm formation, the location inside, and the considered time scales, the material is rather fluid- or solid-like or somewhere in between.
From a physical perspective, such biological systems are described 
as active matter~\cite{marchetti2013hydrodynamics,bechinger2016}. It concerns the collective behavior of active, self-propelled agents, frequently embedded in a surrounding medium.
However, most studies %
refer to one of two categories, depending on the considered long-term flow behavior: active fluids or active solids.

Active fluids lack long-term elastic restoring forces and are capable of terminal flow.
Primary examples are fluid suspensions of microswimmers~\cite{lauga2009hydrodynamics,elgeti2015physics}, such as bacteria or artificial self-propelled particles, and active nematics~\cite{doostmohammadi2018active}.
As described by the seminal Vicsek model~\cite{vicsek1995novel} and its continuum counterpart, the Toner--Tu theory~\cite{toner1998flocks,toner2005hydrodynamics}, collections of active agents capable of unrestricted long-term movement may develop polar orientational order and directed collective motion.
In addition to these flocking or swarming states~\cite{jeckel2019learning,be2020phase}, active fluids exhibit various spatiotemporal patterns~\cite{be2019statistical,nishiguchi2018engineering,aranson2022bacterial,reinken2024pattern} including active turbulence~\cite{dombrowski2004self,wensink2012meso,reinken2018derivation,alert2021active}.

In contrast to active fluids, active solids are unable to support terminal flow due to elastic restoring forces.
Instead, they exhibit reversible elastic deformations.
In theoretical descriptions, the continuous nature of the embedding medium is often reduced to elastic spring networks of connected active agents~\cite{shen2016probing,ferrante2013elasticity,hernandez2024model,baconnier2024noise,laang2024topology}.
Other examples 
are driven by contractile or extensile active elements and resulting stresses within the medium~\cite{hawkins2014stress,maitra2019oriented}, as in muscular tissue~\cite{needleman2017active}.
Recent topics of interest in active solids include the nature of excitations~\cite{baconnier2022selective,caprini2023entropons} and the phenomenon of odd elasticity~\cite{scheibner2020odd,fruchart2023odd}.

There are numerous theoretical studies on either active solid-like~\cite{kopf2013non,maitra2019oriented,scheibner2020odd} or active fluid-like materials with simple~\cite{toner2005hydrodynamics,saintillan2013active,reinken2018derivation} or complex~\cite{hemingway2015active,mathijssen2016upstream,hemingway2016viscoelastic,plan2020active,choudhary2023orientational,reinken2024vortex,reinken2025self} rheologies.
However, a comprehensive exploration of the spatiotemporal dynamics covering these two limiting cases and the broad spectrum in between is yet to be presented.

We aim to bridge this gap and focus on the interplay of viscous and elastic properties in active media on substrates.
To this end, we employ a unified description that can be tuned continuously from the limit of active fluid-like to active solid-like properties.
This framework, which covers a broad range of viscoelastic behavior in between, is derived in a companion article~\cite{reinken2025unified}. It is motivated from a generic microscopic picture of active agents and combined with a memory-based continuum description of the embedding medium. 
Basic, spatially uniform analytical solutions for resulting dynamic states were obtained for small idealized systems. %

Here, we leave this idealization and present a comprehensive dynamical state diagram. %
It reveals fascinating and novel dynamic states 
when varying the activity of the suspension and the rheological properties from fluid\text{-,} via viscoelastic, to solid-like. 
Numerical calculations uncover complex spatiotemporal states, including stripe-like patterns in the active viscoelastic regime and separated oscillating domains for increasingly solid-like systems.
Moreover, we illustrate how tuning the rheological properties in such active suspensions can be exploited for active collective transport on the microscopic scale. This tuning allows for intermediate storage times, maintaining the spatial arrangement within a set of cargo units. 

In short, in the theoretical framework, the behavior is governed by the interplay of three continuum fields~\cite{reinken2025unified}.
The coarse-grained polar order parameter field $\mathbf{P}$ characterizes the local alignment of active units. It couples to the overall velocity field $\mathbf{v}$  
and to the memory field of elastic displacements $\mathbf{u}$. The latter captures the currently memorized deformation history of the film.

First, the evolution of $\mathbf{P}$ in rescaled units follows as
\begin{eqnarray}
\partial_{{t}} \mathbf{P} + {\mathbf{v}} \cdot {\nabla} \mathbf{P} &= &{}- \mathbf{P} + {\nabla}^2 \mathbf{P} + {\boldsymbol{\Omega}} \cdot \mathbf{P}
\nonumber\\ 
&&{}+ {\gamma}_\mathrm{a} {\mathbf{v}} \cdot [(2 + \mathbf{P}\cdot\mathbf{P})\mathbf{I}/3 - \mathbf{P}\mathbf{P}] \, ,
\label{eq:EvolutionEquationPolarOrder}
\end{eqnarray}
where $\mathbf{I}$ is the unit matrix. 
In addition to advection via the flow field $\mathbf{v}$, Eq.~(\ref{eq:EvolutionEquationPolarOrder}) includes reorientation via the vorticity tensor $\boldsymbol{\Omega} = [(\nabla \mathbf{v})^\top - (\nabla \mathbf{v})]$, where $^\top$ denotes the transpose.
The first two terms on the right-hand side arise due to orientational and translational diffusion, while the diffusion coefficients were used to rescale time and space. %
The term proportional to the alignment strength $\gamma_\mathrm{a}$ captures the tendency of active units to align with the overall flow field $\mathbf{v}$ near %
a supporting substrate~\cite{brotto2013hydrodynamics,dadhichi2018origins,maitra2020swimmer,liu2021viscoelastic}.
Similar terms are included in various discretized models of active solids~\cite{szabo2006phase,lam2015self,baconnier2022selective,baconnier2025self}.
Nonlinearity limits the magnitude of the polar order parameter to $|\mathbf{P}|\leq 1$.

Second, the velocity field in the low-Reynolds-number limit is determined via the force balance
\begin{equation}
\label{eq:StokesEquation}
\mathbf{0} = - {\nabla} {p} + {\eta} {\nabla}^2 {\mathbf{v}}  + {\mu}{\nabla}^2 {\mathbf{u}} - {\nu}_\mathrm{v} {\mathbf{v}} - {\nu}_\mathrm{d} {\mathbf{u}} + \nu_\mathrm{p} \mathbf{P} \, .
\end{equation}
We here assume an incompressible material, $\nabla \cdot \mathbf{v} = 0$, enforced by the pressure field $p$. Besides hydrodynamic viscous stress, the equation includes elastic stresses due to previous strain deformations of the material \cite{temmen2000convective, pleiner2004nonlinear, menzel2025linear, reinken2025unified}.
Viscosity $\eta$ and shear modulus $\mu$ together with the elastic memory field $\mathbf{u}$ quantify these viscous and elastic stresses.
Underlying friction with the substrate is set by $\nu_\mathrm{v}$ and slows down movement.
Similarly, elastic restoring forces quantified by $\nu_\mathrm{d}$ result from interactions with the substrate and pull material elements back to their memorized positions.
Active forcing drives the dynamics of the film and is set by $\nu_\mathrm{p}$. 

Third, the dynamics of $\mathbf{u}$ is governed by
\begin{equation}
\label{eq:DisplacementFieldEvolution}
\partial_t \mathbf{u} + \mathbf{v} \cdot \nabla \mathbf{u}  = \mathbf{v} - \tau_\mathrm{d}^{-1} \mathbf{u} - \nabla q\, .
\end{equation}
These dynamics result systematically for elastic strain in established generalized hydrodynamic theories that include elasticity \cite{martin1972unified, temmen2000convective, pleiner2004nonlinear, menzel2025linear, reinken2025unified}. From the hydrodynamic Eulerian perspective, $\mathbf{u}$ contains the memory of the locations from where the material elements have been displaced to their current positions. We therefore rather call it memory field instead of displacement field \cite{puljiz2019memory}. For viscoelastic materials, this memory may fade away over time \cite{temmen2000convective, pleiner2004nonlinear, menzel2025linear, reinken2025unified}.
On the right-hand side of Eq.~(\ref{eq:DisplacementFieldEvolution}) overall motion $\mathbf{v}$ generates displacements and thus associated memory. Conversely, the memory of displacement $\mathbf{u}$ decays with a relaxation time $\tau_\mathrm{d}$ for incomplete elasticity.
This relaxation process implies a net flow.
Full elastic reversibility for solids is found in the limit $\tau_\mathrm{d} \to \infty$.
Finite values $\tau_\mathrm{d}$ describe a broad range of viscoelastic fluids, while the limit $\tau_\mathrm{d}\to0$ refers to purely viscous fluids.
Thus, rheology can be tuned continuously from fluid- to solid-like behavior via $\tau_\mathrm{d}$.
Equation~(\ref{eq:DisplacementFieldEvolution}) includes convective transport $\mathbf{v}\cdot\nabla\mathbf{u}$ by overall flow. In combination with Eq.~(\ref{eq:StokesEquation}), it represents a first step beyond linear elasticity~\cite{puljiz2019memory}.
This description can be derived from a generic continuum theory~\cite{temmen2000convective} based purely on conservation laws and symmetry arguments upon linearization in $\nabla \mathbf{u}$~\cite{reinken2025unified}.
The contribution $-\nabla q$ arises from incompressibility $\nabla \cdot \mathbf{u} = 0$ %
under linearized strains~\cite{reinken2025unified}. %

\begin{figure}
\includegraphics[width=0.999\linewidth]{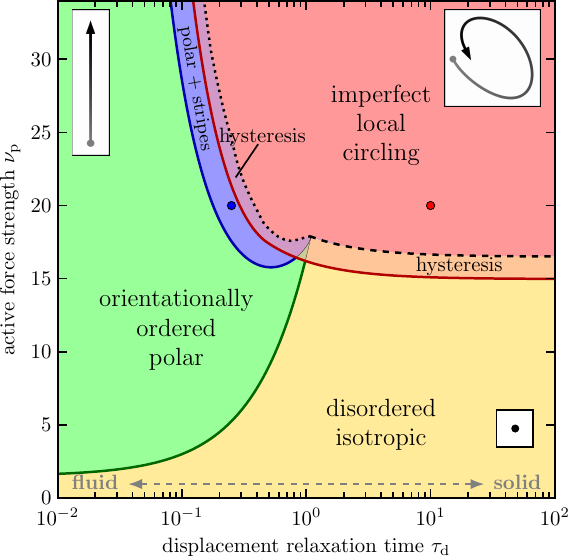}
\caption{\label{fig:StateDiagram}Dynamical state diagram as a function of the displacement relaxation time $\tau_\mathrm{d}$ and strength of active forcing $\nu_\mathrm{p}$. From left to right, %
the influence of elasticity increases. 
Colors indicate regions of different stable dynamical states. %
Insets in the corners illustrate idealized tracer trajectories. %
Blue and red dots refer to the dynamical states explored in more detail in Figs.~\ref{fig:Stripes} and \ref{fig:OscDomains}, respectively. Remaining parameter values are 
$\gamma_\mathrm{a} = 1$, $\eta = 1$, $\mu = 1$, $\nu_\mathrm{v} = 1$, and $\nu_\mathrm{d} = 10$.}
\end{figure}

Previous derivation of the theory suggested three idealized, analytical, spatially uniform mathematical solutions~\cite{reinken2025unified}. We here challenge their stability, amended by numerical simulations. For technical details see the Supplemental Material~\cite{suppl}, which includes Refs.~\onlinecite{canuto2007spectral,durran2010numerical,harris2020array}. The central goal is to construct the dynamical state diagram for active films of viscoelastic rheology. 

Figure~\ref{fig:StateDiagram} shows a corresponding illustration. %
There, from bottom to top, %
we increase the strength of active forcing $\nu_\mathrm{p}$, starting from passive systems ($\nu_\mathrm{p} = 0$).
From left to right, with increasing displacement relaxation time $\tau_\mathrm{d}$, we modify the system from rather viscous, fluid-like behavior towards an increasingly elastic, solid-like system.

Indeed, on the left of Fig.~\ref{fig:StateDiagram}, for $\tau_\mathrm{d}\rightarrow0$ and fluid-like systems, we recover the generic stationary states. At low activity, this is a macroscopically disordered, isotropic state ($\mathbf{P}=\mathbf{0}=\mathbf{v}$). 
With increasing activity, it transitions to an orientationally ordered polar state ($\mathbf{P}\neq\mathbf{0}\neq\mathbf{v}$) of straight motion (see the inset on the top left). Both states in the marked regimes are linearly stable~\cite{suppl}. 
Corresponding phenomenologies are well known, %
for example, from the Vicsek model~\cite{vicsek1995novel}, the Toner--Tu theory~\cite{toner1998flocks,toner2005hydrodynamics}, and related extensions~\cite{wensink2012meso,reinken2024pattern,alert2021active}.

Raising the influence of elasticity by increasing $\tau_\mathrm{d}$, we find that both states extend to the right in Fig.~\ref{fig:StateDiagram}. Yet, elasticity seems to hinder motion. Illustratively, the elastic restoring forces pin the active agents. Therefore, the phase boundary between polar moving (green) and isotropic resting (yellow) areas turns towards higher activity with increasing elasticity. The phase boundary results from linear stability analysis and is confirmed numerically~\cite{suppl}.

\begin{figure}
\includegraphics[width=0.999\linewidth]{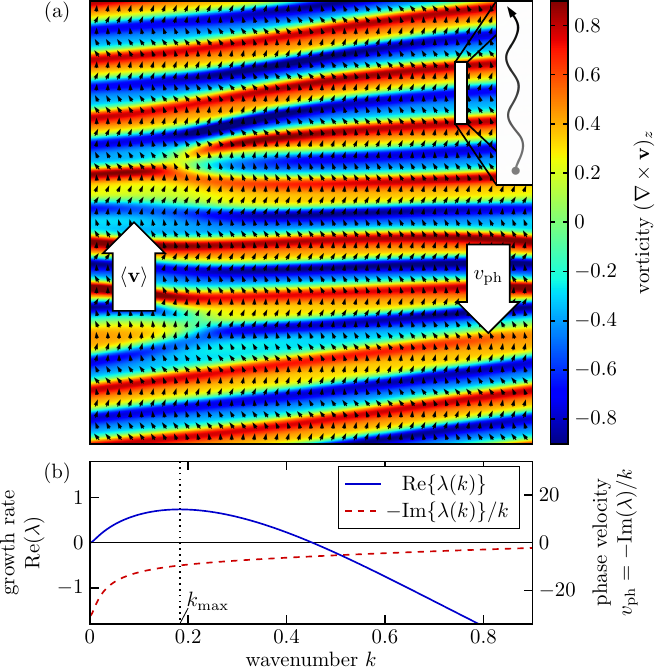}
\caption{\label{fig:Stripes}Stripe pattern associated with the blue region in Fig.~\ref{fig:StateDiagram}.  
(a) Snapshot of the vorticity field $\omega=(\nabla\times\mathbf{v})_z$ in a system of size $200\times 200$ at $\tau_\mathrm{d} = 0.25$, see the blue dot in Fig.~\ref{fig:StateDiagram}.
Small dark arrows indicate the velocity field $\mathbf{v}$. 
Transport of the patterns, given by the phase velocity $v_\mathrm{ph}=-\mathrm{Im}\{\lambda(k)\}/k$, is opposite to the averaged active flow direction  $\langle\mathbf{v}\rangle$. 
Associated material trajectories are snake-like, see the top right inset. 
(b) Largest growth rates $\mathrm{Re}\{\lambda(k)\}$ of modes $\mathbf{k}\|\langle\mathbf{v}\rangle$ when starting from a uniform polar state, and phase velocity $v_\mathrm{ph}=-\mathrm{Im}\{\lambda(k)\}/k$. 
Remaining parameter values are $\gamma_\mathrm{a} = 1$, $\eta = 1$, $\mu = 1$, $\nu_\mathrm{v} = 1$, $\nu_\mathrm{d} = 10$, and $\nu_\mathrm{p} = 20$.}
\end{figure}

The picture changes notably when we increase elasticity at elevated levels of activity from the polar state. Then novel, qualitatively different behavior emerges. It is marked by the blue region in Fig.~\ref{fig:StateDiagram}. Numerical evaluations indicate a stripe-like dynamic state, see Fig.~\ref{fig:Stripes}(a). The inset depicts associated wiggling, snake-like motion. 

Indeed, the state of globally uniform polar orientational order becomes linearly unstable in this region~\cite{suppl}. 
For the associated linear stability analysis \cite{suppl}, we summarize all dynamic quantities in a vector $\mathbf{V}$. We consider small perturbations $\delta\mathbf{V}$ in terms of Fourier modes of wavevector $\mathbf{k}$, $\delta\mathbf{V} = \delta\hat{\mathbf{V}} e^{\lambda t} e^{i \mathbf{k} \cdot \mathbf{x}}$. 
The growth rate $\mathrm{Re}\{\lambda(k=|\mathbf{k}|)\}$ is plotted in Fig.~\ref{fig:Stripes}(b),
with $\mathbf{k}$ oriented parallel to $\mathbf{P}$. 
Importantly, the fastest-growing mode is of finite wavenumber $k_\mathrm{max} > 0$.
Thus, the instability features a certain length scale $\Lambda = 2 \pi /k_\mathrm{max}$, at least close to its onset.
This length is reflected by spatial modulations along the migration direction, see Fig.~\ref{fig:Stripes}(a). 
There, the local velocity field is indicated by the dark arrows. Its directions deviate periodically in space from the average migration direction given by $\langle\mathbf{v}\rangle$.  
The periodic modulation is further visualized by the vorticity field $\omega = (\nabla \times \mathbf{v})_z$, see the colored stripe pattern in Fig.~\ref{fig:Stripes}(a). 

Moreover, from $\mathrm{Im}\{\lambda(k)\}$, we calculate the phase velocity of the pattern, $v_\mathrm{ph}=-\mathrm{Im}(\lambda)/k$. It characterizes the traveling speed of the individual waves.
Interestingly, we always find $v_\mathrm{ph}<0$. %
Consequently, the emerging structures travel in the direction opposite to global motion $\langle\mathbf{v}\rangle$, %
see Fig.~\ref{fig:Stripes}(a), see also Supplemental Movie~1. 

Banded structures are observed in various systems of active matter.
Examples include traveling density bands in systems of self-propelled particles~\cite{caussin2014emergent}, bands in suspensions of microswimmers subject to external fields~\cite{reinken2019anisotropic}, as well as active nematic systems~\cite{patelli2019understanding}.
In contrast to these earlier observations, the traveling band-like structures we find in our study at the boundary of the ordered polar state are induced by elasticity, only observable in the viscoelastic regime.

\begin{figure*}
\includegraphics[width=0.999\linewidth]{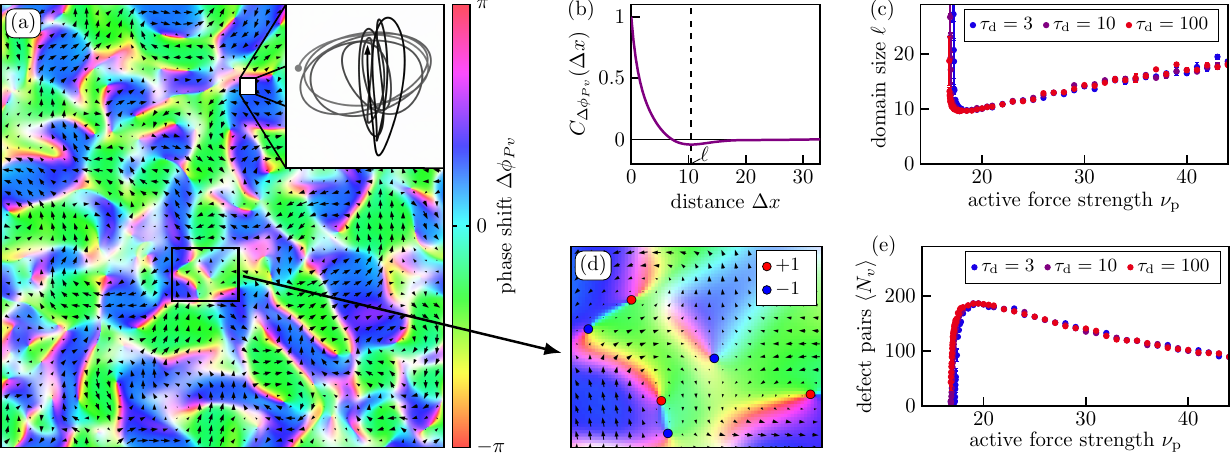}
\caption{\label{fig:OscDomains}Heterogeneous state of imperfect local circling motion. 
(a) At each spot, the vectors of polar orientational order $\mathbf{P}$ and collective flow $\mathbf{v}$ (dark arrows) rotate over time. 
Domains are identified by the field of phase shift $\Delta \phi_{Pv}$ (color bar) between $\mathbf{P}$ and $\mathbf{v}$. 
Positive and negative values are connected to clockwise and counterclockwise rotation, respectively. 
A sample trajectory is shown in the top right corner.
The size of the snapshot is $100\times 100$.
(b) Averaged spatial correlation function $C_{\Delta \phi_{Pv}}(\Delta x)$ of the phase shift $\Delta \phi_{Pv}$ between any two points of evaluation with distance $\Delta x$ from each other. 
The associated characteristic length scale $\ell$ of the dynamic patterns 
is identified with the first minimum of $C_{\Delta \phi_{Pv}}(\Delta x)$. 
(c) Typical domain size $\ell$ as a function of active force strength $\nu_\mathrm{p}$ for different values of the relaxation time $\tau_\mathrm{d}$.
(d) Enlarged view of the marked area in (a) showing $+1$ and $-1$ defects in the velocity field as red and blue dots.
(e) Time-averaged number of defect pairs $\langle N_v\rangle$ in the velocity field of a system of size $200\times 200$ as a function of $\nu_\mathrm{p}$.
Remaining parameter values are $\gamma_\mathrm{a} = 1$, $\eta = 1$, $\mu = 1$, $\nu_\mathrm{v} = 1$, $\nu_\mathrm{d} = 10$, and $\nu_\mathrm{p} = 20$.}
\end{figure*}

Further increasing the influence of elasticity via $\tau_\mathrm{d}$ to the right in Fig.~\ref{fig:StateDiagram}, we cross the red line. It marks the existence of a basic analytical solution, corresponding to rotating spatially uniform polar order $\mathbf{P}$, flow $\mathbf{v}$, and elastic memory $\mathbf{u}$~\cite{suppl}. Yet, we do not observe this state in our spatially extended numerical systems.  

Here, numerically, we identify a hysteretic regime until we reach the dotted line in Fig.~\ref{fig:StateDiagram}~\cite{suppl}. 
In this area, the emergent state depends sensitively on the initial conditions. 
On the one hand, upon initialization by a weakly perturbed state of global polar orientational order, still the stripe patterns develop. 
On the other hand, upon initialization by a weak perturbation of the above-mentioned rotational analytical solution~\cite{suppl}, rotations of the three fields are maintained. 
Further to the right of the dotted line in Fig.~\ref{fig:StateDiagram}, the stripes are not found as a final dynamic state any more. 
Although transients can be observed for different initializations, at the end, we find a state of locally rotating fields $\mathbf{P}$, $\mathbf{v}$, and $\mathbf{u}$, see Fig.~\ref{fig:OscDomains}(a). 
In fact, this state is not spatially uniform. The system splits up into different domains of rotation. Thus, it becomes spatially heterogeneous. 

Physically, the elastic restoring forces for elevated $\tau_\mathrm{d}$ become too strong to allow for persistent motion. The globally ordered polar or stripe-like dynamic states are associated with continuous directed transport. Thus, they are suppressed by higher levels of elasticity. %
Instead, local rotations of the polar order $\mathbf{P}$, flow field $\mathbf{v}$, and memory field $\mathbf{u}$ combine active transport with local elastic anchoring. Imperfect circular trajectories arise, see the top right insets in Figs.~\ref{fig:StateDiagram} and \ref{fig:OscDomains}(a).

The rotations of the fields $\mathbf{P}$, $\mathbf{v}$, and $\mathbf{u}$ are associated with local phase shifts between them. 
These phase shifts can be used to identify the local domains of rotation, see Fig.~\ref{fig:OscDomains}(a). %
For example, the field of phase shift $\Delta \phi_{Pv} = \phi_P - \phi_v$ refers to the local angle between $\mathbf{P}$ and  $\mathbf{v}$.
Remarkably, rotations in different domains may occur with opposite sense of rotation, which is indicated by the different signs in $\Delta \phi_{Pv}$.
Supplemental Movie~2 provides a dynamic visualization of the dynamic state associated with the snapshot in Fig.~\ref{fig:OscDomains}(a).  
From there, it also becomes evident that the domains of correlated phases are constantly evolving in time.

To quantify the length scales of correlated domains, we calculate the equal-time spatial correlation function
\begin{equation}
\label{eq:CorrelationsAngleDifference}
C_{\Delta \phi_{Pv}}(\Delta x) = \big\langle \Delta \phi_{Pv}(\mathbf{x},t)\, \Delta \phi_{Pv}(\mathbf{x} + \hat{\mathbf{n}}\Delta x,t) \big\rangle_{\hat{\mathbf{n}},\mathbf{x},t}\, .
\end{equation}
Here, $\Delta x$ is the distance between any two points in space and $\hat{\mathbf{n}}$ is the direction of their separation. 
$\langle \dots \rangle_{\hat{\mathbf{n}},\mathbf{x},t}$ denotes the average over orientations $\hat{\mathbf{n}}$, space positions $\mathbf{x}$, and time $t$.
For the dynamically evolving state depicted in Fig.~\ref{fig:OscDomains}(a), we plot $C_{\Delta \phi_{Pv}}(\Delta x)$ in Fig.~\ref{fig:OscDomains}(b).
The distinct length scale $\ell$ quantifying the typical domain size is 
identified with the location of the first minimum of $C_{\Delta \phi_{pv}}(\Delta x)$.
Indeed, we find that the length scale $\ell$ of the typical size of the correlated domains is of the same order as the thickness of the stripes $\Lambda/2$ referred to in Fig.~\ref{fig:Stripes}(a).
Figure.~\ref{fig:OscDomains}(c) shows $\ell$ as a function of the strength of active forcing $\nu_\mathrm{p}$ for different values of the relaxation time $\tau_\mathrm{d}$ within the red region of Fig.~\ref{fig:StateDiagram}. 
Mainly, we find that $\ell$ obtains a minimum, %
but then increases continuously with increasing activity $\nu_\mathrm{p}$. Thus, the structures become increasingly correlated. Approaching the transition to the isotropic state, see below, $\ell$ diverges and the structures are highly correlated in space.
Overall, we find that the relaxation time of elasticity in terms of $\tau_\mathrm{d}$ has only little influence on $\ell$ for the considered parameters.

As for various other systems of active matter~\cite{giomi2013defect,doostmohammadi2018active,alert2021active,rana2024defect}, defects in the velocity and polar order parameter fields are continuously pairwise created and annihilated. Their mobility is visualized in Supplemental Movie 3.
In contrast to active nematics~\cite{giomi2013defect,doostmohammadi2018active}, we observe defects of integer value in our polar system.
$+1$ defects correspond to vortices, whereas $-1$ defects locate (saddle) points in between, see Fig.~\ref{fig:OscDomains}(d). We expect the number of defects to behave oppositely to the domain size $\ell$ in Fig.~\ref{fig:OscDomains}(c). Indeed, 
Fig.~\ref{fig:OscDomains}(e) confirms this relation for the time-averaged number of defect pairs $\langle N_v\rangle$ in the velocity field as a function of activity $\nu_\mathrm{p}$ for different values of $\tau_\mathrm{d}$.
The time-averaged number of defect pairs for the field $\mathbf{P}$ yields qualitatively the same result~\cite{suppl}.

Finally, we address the transition across the lower boundary of the red region in Fig.~\ref{fig:StateDiagram}.
Increasing the strength of activity $\nu_\mathrm{p}$ from the bottom, the isotropic state only becomes linearly unstable at the dashed black line~\cite{suppl}. Indeed, we observe this state numerically up to the dashed line for weakly perturbed isotropic initialization. Conversely, dynamic states of circling %
are found down to the red line, if the system is initialized by the weakly perturbed idealized rotational analytical solution~\cite{suppl}. Again, this scenario reveals a region of hysteresis. 
A subcritical bifurcation is identified in this area in a simplified analytical approach, as discussed separately for the idealized, spatially uniform analytical solutions~\cite{reinken2025unified}.

\begin{figure}
\includegraphics[width=0.999\linewidth]{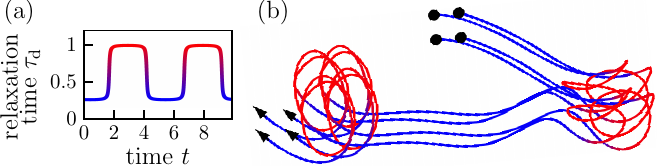}
\caption{\label{fig:Trajectories}Rheologically controlled transport of an array of cargo objects. (a) We periodically vary the relaxation time $\tau_\mathrm{d}$ of elastic displacements between $0.25$ and $1$. Thus, we tune the degree of elasticity. (b) Resulting trajectories for an array of four transported cargo objects. Tuning rheology allows in a controlled way to switch between persistent collective transport and intermediate storage times, roughly maintaining the spatial order within the transported cargo array. Blue and red parts in (a) and (b) correspond to each other and to the blue and red regions in the state diagram in Fig.~\ref{fig:StateDiagram}. %
The remaining parameters are 
$\gamma_\mathrm{a} = 1$, $\eta = 1$, $\mu = 1$, $\nu_\mathrm{v} = 1$, $\nu_\mathrm{d} = 10$, and $\nu_\mathrm{p} = 20$. The initial edges in the cargo array are of length $1$.}
\end{figure}

As an immediate perspective, changing the rheology of active films allows to %
control collective transport.
For example, light and external electric fields enable remote dynamic tuning of the rheology of photorheological~\cite{oh2013simple,yang2014multistimuli,tomatsu2011photoresponsive} or electrorheological~\cite{kuznetsov2022electrorheological} materials. Exploiting such effects facilitates externally controlled transport of cargo objects by active carrier media. Persistent polar orientational order implies directed transport, while switching to the nonpolar states facilitates intermediate storage times. Elevated activity implies increased speed of transport. It requires intermediate switching to the state of imperfect local circling during rest intervals. 
We numerically demonstrate this scenario in Fig.~\ref{fig:Trajectories} by periodically changing the relaxation time $\tau_\mathrm{d}$, see Fig.~\ref{fig:Trajectories}(a). %
Resulting trajectories in Fig.~\ref{fig:Trajectories}(b) show the transport of an array of cargo objects, %
including intermediate storage periods, %
while maintaining the internal positional order within the array. Suitable preparation of the substrate can guide the transport. 
The approach opens new routes in microfluidics powered by the activity of microswimmers such as bacteria~\cite{darnton2004moving,kaehr2009high,kim2008microfluidic,thampi2016active}.
Instead varying the relaxation time in space 
suggests a potential mechanism for persistent particle trapping.

In summary, we uncover a rich world of dynamic states that result from the interplay between viscoelasticity and activity. %
We construct the overall dynamical state diagram
and observe the emergence of novel dynamic states. First, the stripe-like patterns are neither found for purely viscous active fluids nor for purely elastic active solids. Their origin is thus different from earlier observed banded structures~\cite{caussin2014emergent,patelli2019understanding,reinken2019anisotropic}.
Second, the heterogeneous states of imperfect circling result from elastic forces within the system, in contrast to heterogeneous states observed in purely viscous suspensions of microswimmers~\cite{alert2021active,rana2024defect}.

Besides suggesting engineered active systems to control directed transport of cargo on the microscale, our description relates to the complex dynamical behavior in living biological matter. 
Biofilms and similar collections of cells on substrates show complex flow patterns~\cite{yashunsky2022chiral,xu2023autonomous}, which are reminiscent of the structures observed in Fig.~\ref{fig:OscDomains}(a).
Bacterial biofilms~\cite{hall2004bacterial,worlitzer2022biophysical,jana2020nonlinear}, in particular, may change their rheological characteristics dynamically and continuously from purely viscous via viscoelastic to elastic as a function of time and position within the film.

Generally, we wish to stimulate by our work 
experimental and theoretical research on biological and artificial active materials that exploit dynamical changes in their rheological properties to gain functionality. 
In this sense, future research may address the further development of adaptive living materials~\cite{an2022engineered}, active synthetic systems~\cite{kim2013soft,wu2020medical}, and actuators~\cite{kim2022magnetic,li2022soft}.
Our results demonstrate the potential of tunable viscoelasticity to accommodate varying needs and demand during application. 

\begin{acknowledgments}
We thank the Deutsche Forschungsgemeinschaft (German Research Foundation, DFG) for support through the Research Grant no.\ ME 3571/5-1. Moreover, A.M.M. acknowledges support through the Heisenberg Grant no.\ ME 3571/4-1. 
\end{acknowledgments}

The numerical code used to generate the data in support of the findings of this article is openly available~\cite{code2025rheologically}. 


%

\end{document}


\title{Rheologically tuned modes of collective transport in active viscoelastic films\\[1\baselineskip] \textit{Supplemental Material}}

\author{Henning Reinken}
\email{henning.reinken@ovgu.de}
\affiliation{Institut f\"ur Physik, Otto-von-Guericke-Universität Magdeburg, Universitätsplatz 2, 39106 Magdeburg, Germany} 

\author{Andreas M. Menzel}
\email{a.menzel@ovgu.de}
\affiliation{Institut f\"ur Physik, Otto-von-Guericke-Universität Magdeburg, Universitätsplatz 2, 39106 Magdeburg, Germany} 

\begin{abstract}
We here describe the linear stability analysis that forms the basis for the dynamical state diagram in the main article. In this context, we also overview a known analytical, spatially uniform, rotational solution of the equations of motion. Additionally, we provide information on the numerical methods used to find our results and the associated Supplemental Movies. Finally, we discuss topological defects and include details on the periodic variation of the rheological properties that enable the switching between modes of collective transport as presented in the main text.
\end{abstract}

\date{\today}

\maketitle

\section{Linear stability analysis}
\label{app:LinearStabilityAnalysis}

The dynamical state diagram presented in the main article is obtained on the basis of a stability analysis of the spatially uniform stationary solutions.
When the theoretical description was derived, two basic, stationary, spatially uniform solutions were identified \cite{reinken2025unified}.
Among them are an isotropic, quiescent state ($\mathbf{P}=\mathbf{v}=\mathbf{u}=\mathbf{0}$) and a polar state of directed motion ($\mathbf{P}\neq\mathbf{0}$, $\mathbf{v}\neq\mathbf{0}$, $\mathbf{u}\neq\mathbf{0}$).
Inferring their linear stability %
from analytical considerations is straightforward and presented in the following.

As a starting point, we add small perturbations to the respective stationary solution,
\begin{equation}
\label{eq:Perturbations}
\begin{aligned}
\begin{pmatrix} P_x \\ P_y \end{pmatrix} &= \begin{pmatrix} P_0 \\ 0 \end{pmatrix} + \begin{pmatrix} \delta P_x \\ \delta P_y \end{pmatrix} , \\
\begin{pmatrix} v_x \\ v_y \end{pmatrix} &= \begin{pmatrix} v_0 \\ 0 \end{pmatrix} + \begin{pmatrix} \delta v_x \\ \delta v_y \end{pmatrix} , \\
\begin{pmatrix} u_x \\ u_y \end{pmatrix} &= \begin{pmatrix} u_0 \\ 0 \end{pmatrix} + \begin{pmatrix} \delta u_x \\ \delta u_y \end{pmatrix} , \\
p &= p_0 + \delta p\, ,\\
q &= q_0 + \delta q\, .\
\end{aligned}
\end{equation}
Here, $P_0$, $v_0$, and $u_0$ are set to zero in the case of the trivial, isotropic solution.
The stationary, uniform, polar solution is given by~\cite{reinken2025unified}
\begin{equation}
\label{eq:PolarSolution}
\begin{aligned}
P_0 &= \sqrt{1 - \frac{3(\nu_\mathrm{v} + \nu_\mathrm{d} \tau_\mathrm{d})}{2 \gamma_\mathrm{a} \nu_\mathrm{p}}}\, , \\
v_0 &= \frac{\nu_\mathrm{p} }{\nu_\mathrm{v} + \nu_\mathrm{d} \tau_\mathrm{d}} P_0 \, , \\
u_0 &= \frac{\nu_\mathrm{p} \tau_\mathrm{d}}{\nu_\mathrm{v} + \nu_\mathrm{d} \tau_\mathrm{d}} P_0 \, .
\end{aligned}
\end{equation}
Perturbations are introduced in terms of Fourier modes of wavevector $\mathbf{k} = (k_x,k_y)$,
\begin{equation}
\label{eq:PerturbationFourierMode}
\delta\mathbf{V}=
\begin{pmatrix} \delta P_x \\ \delta P_y \\ \delta v_x \\ \delta v_y \\ \delta u_x \\ \delta u_y \\ \delta p \\ \delta q \end{pmatrix} = \begin{pmatrix} \delta \hat{P}_x \\ \delta \hat{P}_y \\ \delta \hat{v}_x \\ \delta \hat{v}_y \\ \delta \hat{u}_x \\ \delta \hat{u}_y \\ \delta \hat{p} \\ \delta \hat{q} \end{pmatrix} e^{\lambda t} e^{i \mathbf{k} \cdot \mathbf{x}}\, ,
\end{equation}
where $\delta \hat{P}_x$, $\delta \hat{P}_y$, $\delta \hat{v}_x$, $\delta \hat{v}_y$, $\delta \hat{u}_x$, $\delta \hat{u}_y$, $\delta \hat{p}$, and $\delta \hat{q}$ are the perturbation amplitudes.
The aim of the stability analysis is to determine the complex quantity $\lambda$ as a function of the wavevector $\mathbf{k}$ of the perturbation and the material parameters.
Then, the real part of $\lambda$ determines the growth rate, signifying an instability at wavevector $\mathbf{k}$ if $\mathrm{Re}\{\lambda(\mathbf{k})\} > 0$.
If the imaginary part of $\lambda$ is nonzero as well, the instability is oscillatory, for example leading to traveling waves.
In that case, the phase velocity $v_\mathrm{ph}$ is calculated as 
\begin{equation}
\label{eq:PhaseVelocity}
v_\mathrm{ph}(\mathbf{k}) = {}- \mathrm{Im}\{\lambda(\mathbf{k})\}/|\mathbf{k}|\, 
\end{equation}
and determines the traveling speed of the individual waves as a function of the wavevector.

To continue, we insert Eqs.~(\ref{eq:Perturbations}) and (\ref{eq:PerturbationFourierMode}) into  Eqs.~(1)--(3) of the main text.
Keeping only terms linear in the perturbation,
this procedure yields
\begin{equation}
\begin{aligned}
\lambda \delta \hat{P}_x = &- v_0 i k_x \delta \hat{P}_x - |\mathbf{k}|^2 \delta \hat{P}_x -  P_0 -  \delta \hat{P}_x + \kappa i k_x \delta \hat{v}_x P_0\\
&+ \frac{2 \gamma_\mathrm{a}}{3} v_0 + \frac{2 \gamma_\mathrm{a}}{3} \delta \hat{v}_x - \frac{2\gamma_\mathrm{a}}{3} (v_0 P_0^2 + 2 P_0 v_0 \delta \hat{P}_x + P_0^2 \delta \hat{v}_x ) \, ,\\ 
\lambda \delta \hat{P}_y = &- v_0 i k_x \delta \hat{P}_y - |\mathbf{k}|^2 \delta \hat{P}_y -  \delta \hat{P}_y + \frac{1}{2} i k_x \delta \hat{v}_y P_0  - \frac{1}{2} i k_y \delta \hat{v}_x P_0 \\
&+ \kappa \frac{1}{2} i k_x \delta \hat{v}_y P_0 + \kappa \frac{1}{2} i k_y \delta \hat{v}_x P_0 + \frac{2\gamma_\mathrm{a}}{3} \delta \hat{v}_y + \frac{\gamma_\mathrm{a}}{3} P_0^2 \delta \hat{v}_y - \gamma_\mathrm{a} P_0 v_0 \delta \hat{P}_y\,,
\end{aligned}
\end{equation}
\begin{equation}
\begin{aligned}
0 = &- i k_x \delta \hat{p} - \eta |\mathbf{k}|^2 \delta \hat{v}_x - \nu_\mathrm{v} v_0 - \nu_\mathrm{v} \delta \hat{v}_x - \mu |\mathbf{k}|^2 \delta \hat{u}_x - \nu_\mathrm{d} u_0 - \nu_\mathrm{d} \delta \hat{u}_x + \nu_\mathrm{p} P_0 + \nu_\mathrm{p} \delta \hat{P}_x \, , \\
0 = &- i k_y \delta \hat{p} - \eta |\mathbf{k}|^2 \delta \hat{v}_y - \nu_\mathrm{v} \delta \hat{v}_y - \mu |\mathbf{k}|^2 \delta \hat{u}_y - \nu_\mathrm{d} \delta \hat{u}_y  + \nu_\mathrm{p} \delta \hat{P}_y \, ,
\end{aligned}
\end{equation}
\begin{equation}
\begin{aligned}
\lambda \delta \hat{u}_x = &- v_0 i k_x\delta \hat{u}_x + v_0 + \delta \hat{v}_x - \tau_\mathrm{d}^{-1} u_0 - \tau_\mathrm{d}^{-1} \delta \hat{u}_x - i k_x \delta \hat{q}\, , \\
\lambda \delta \hat{u}_y = &- v_0 i k_x\delta \hat{u}_y + \delta \hat{v}_y - \tau_\mathrm{d}^{-1} \delta \hat{u}_y - i k_y \delta \hat{q}\, , 
\end{aligned}
\end{equation}
\begin{equation}
\begin{aligned}
k_x \delta \hat{v}_x + k_y \delta \hat{v}_y = 0\, , \qquad k_x \delta \hat{u}_x + k_y \delta \hat{u}_y = 0\, .
\label{eq:incomp_k}
\end{aligned}
\end{equation}
All terms not containing any perturbation vanish when inserting one of the two spatially uniform solutions mentioned above~\cite{reinken2025unified}.
We further find that the perturbation of the variable $q$ must vanish, $\delta \hat{q} = 0$. %

The equations can then be rewritten in the form
\begin{equation}
\label{eq:LSAPMatrixForm}
\lambda \delta\hat{\mathbf{P}} = \mathbf{M}_{PP} \cdot \delta \hat{\mathbf{P}} + \mathbf{M}_{Pv} \cdot \delta \hat{\mathbf{v}} \, ,
\end{equation}
\begin{equation}
\label{eq:LSAvMatrixForm}
0 = -(\eta |\mathbf{k}|^2 + \nu_\mathrm{v}) \delta \hat{\mathbf{v}} - (\mu |\mathbf{k}|^2 + \nu_\mathrm{d}) \delta \hat{\mathbf{u}}  + \nu_\mathrm{p} \delta \hat{\mathbf{P}} - i \mathbf{k} \delta \hat{p} \, ,
\end{equation}
\begin{equation}
\label{eq:LSAuMatrixForm}
\lambda \delta\hat{\mathbf{u}} = - (v_0 i k_x +\tau_\mathrm{d}^{-1}) \delta \hat{\mathbf{u}} + \delta \hat{\mathbf{v}}\, ,
\end{equation}
where the matrices are given by
\begin{equation}
\setlength\arraycolsep{5pt}
\label{eq:LSAMatrixPP}
\mathbf{M}_{PP} = \begin{pmatrix}
-v_0 i k_x - |\mathbf{k}|^2 - 1 - 4\gamma_\mathrm{a} P_0 v_0 /3 & 0 \\[3pt]
0 & -v_0 i k_x - |\mathbf{k}|^2 - 1 - \gamma_\mathrm{a} P_0 v_0
\end{pmatrix},
\end{equation}
\begin{equation}
\setlength\arraycolsep{5pt}
\label{eq:LSAMatrixPv}
\mathbf{M}_{Pv} = \begin{pmatrix}
\kappa i k_x P_0 + 2\gamma_\mathrm{a} /3 - 2\gamma_\mathrm{a} P_0^2/3 & 0 \\[3pt]
- i k_y P_0 /2 + \kappa i k_y P_0 /2 & i k_x P_0 /2 + \kappa ik_x P_0 /2 + 2 \gamma_\mathrm{a} /3 + \gamma_\mathrm{a} P_0^2 /3
\end{pmatrix}.
\end{equation}

To continue, we first eliminate the perturbation of $p$ utilizing the incompressibility conditions.
To this end, we multiply Eq.~(\ref{eq:LSAvMatrixForm}) by $\mathbf{k}$ and use $\mathbf{k} \cdot \delta \hat{\mathbf{v}} = 0$ and $\mathbf{k} \cdot \delta \hat{\mathbf{u}} = 0$, see Eqs.~(\ref{eq:incomp_k}), which yields
\begin{equation}
\label{eq:LSAvPressure1}
0 =  \nu_\mathrm{p} \mathbf{k} \cdot \delta \hat{\mathbf{P}} - i |\mathbf{k}|^2 \delta \hat{p} \, .
\end{equation}
The perturbation $\delta \hat{p}$ is thus determined via
\begin{equation}
\label{eq:LSAvPressure2}
\delta \hat{p} = - i \nu_\mathrm{p} \frac{\mathbf{k}}{|\mathbf{k}|^2} \cdot \delta \hat{\mathbf{P}}\, .
\end{equation}
Inserting this expression into Eq.~(\ref{eq:LSAvMatrixForm}), we obtain
\begin{equation}
\label{eq:LSAvMatrixForm2}
\mathbf{0} = -(\eta |\mathbf{k}|^2 + \nu_\mathrm{v}) \delta \hat{\mathbf{v}} - (\mu |\mathbf{k}|^2 + \nu_\mathrm{d}) \delta \hat{\mathbf{u}}  + \nu_\mathrm{p} (\mathbf{I} - \mathbf{k}\mathbf{k}/|\mathbf{k}|^2) \cdot \delta \hat{\mathbf{P}} \, . \end{equation}
We solve this equation for $\delta \hat{\mathbf{v}}$,
\begin{equation}
\label{eq:LSAvMatrixForm3}
\delta \hat{\mathbf{v}} = -\frac{\mu |\mathbf{k}|^2 + \nu_\mathrm{d}}{\eta |\mathbf{k}|^2 + \nu_\mathrm{v}} \delta \hat{\mathbf{u}}  + \nu_\mathrm{p} \frac{\mathbf{I} - \mathbf{k}\mathbf{k}/|\mathbf{k}|^2}{\eta |\mathbf{k}|^2 + \nu_\mathrm{v}}\cdot \delta \hat{\mathbf{P}} \, . 
\end{equation}
Using this expression to eliminate from Eqs.~(\ref{eq:LSAPMatrixForm}) and (\ref{eq:LSAuMatrixForm}) the perturbation in velocity, we finally find
\begin{equation}
\lambda \delta\hat{\mathbf{P}} = \bigg( \mathbf{M}_{PP} + \nu_\mathrm{p} \mathbf{M}_{Pv} \cdot \frac{\mathbf{I} - \mathbf{k}\mathbf{k}/|\mathbf{k}|^2}{\eta |\mathbf{k}|^2 + \nu_\mathrm{v}}  \bigg) \cdot \delta \hat{\mathbf{P}} -\frac{\mu |\mathbf{k}|^2 + \nu_\mathrm{d}}{\eta |\mathbf{k}|^2 + \nu_\mathrm{v}} \mathbf{M}_{Pv} \cdot \delta \hat{\mathbf{u}} \, ,
\end{equation}
\begin{equation}
\label{eq:LSAuMatrixForm3}
\lambda \delta\hat{\mathbf{u}} = \nu_\mathrm{p}\frac{\mathbf{I} - \mathbf{k}\mathbf{k}/|\mathbf{k}|^2}{\eta |\mathbf{k}|^2 + \nu_\mathrm{v}} \cdot \delta \hat{\mathbf{P}}- \bigg[(v_0 i k_x +\tau_\mathrm{d}^{-1}) \mathbf{I} + \frac{\mu |\mathbf{k}|^2 + \nu_\mathrm{d}}{\eta |\mathbf{k}|^2 + \nu_\mathrm{v}}( \mathbf{I} -  \mathbf{k}\mathbf{k}/|\mathbf{k}|^2)\bigg] \cdot \delta \hat{\mathbf{u}} \, .
\end{equation}

To simplify notation, we summarize all remaining perturbations in a single vector, $\delta \hat{\mathbf{V}}_\mathrm{rem} = (\delta \hat{P}_x,\delta \hat{P}_y,\delta \hat{u}_x,\delta \hat{u}_y)$.
Thus, the set of equations can now be written as the eigenvalue problem
\begin{equation}
\lambda \delta \hat{\mathbf{V}}_\mathrm{rem} = \mathbf{M}\cdot \delta \hat{\mathbf{V}}_\mathrm{rem}\, .
\end{equation}
The complex growth rate $\lambda$ as a function of $\mathbf{k}$ is then identified as the eigenvalues of the $4\times 4$ matrix $\mathbf{M}$.
Here, we use the Python library \textit{NumPy}~\cite{harris2020array} to efficiently evaluate the matrix $\mathbf{M}$ for different parameter values and determine its eigenvalues.
The eigenvalue with the largest real part decides on the stability of the stationary solution under investigation (isotropic or polar).
This result is used in the construction of the state diagram in Fig.~1.
An example of the growth rate and phase velocity given by the eigenvalue with largest real part for the stationary polar solution is shown in Fig.~2(d).

\section{Analytical rotating solution}

In addition to the stationary spatially uniform solutions studied above, Ref.~\onlinecite{reinken2025unified} showed the existence of a basic analytical, spatially uniform solution that is not stationary.
It is characterized by nonvanishing fields of polar orientational order, flow velocity, and memorized elastic displacement ($\mathbf{P}\neq\mathbf{0}$, $\mathbf{v}\neq\mathbf{0}$, $\mathbf{u}\neq\mathbf{0}$). However, these fields continuously rotate according to
\begin{equation}
\label{eq:PolarForm}
\mathbf{P} = P_0 \! \begin{pmatrix} \cos(\omega_0 t) \\ \sin(\omega_0 t) \end{pmatrix}\! , \ \mathbf{v} = v_0\! \begin{pmatrix} \cos(\omega_0 t - \Delta\phi_{Pv}) \\ \sin(\omega_0 t - \Delta\phi_{Pv}) \end{pmatrix}\! , \ \mathbf{u} = u_0 \! \begin{pmatrix} \cos(\omega_0 t - \Delta\phi_{Pu}) \\ \sin(\omega_0 t - \Delta\phi_{Pu}) \end{pmatrix}\! ,
\end{equation}
where $P_0$, $v_0$, and $u_0$ are the magnitudes of $\mathbf{P}$, $\mathbf{v}$, and $\mathbf{u}$, respectively.
The angular frequency is denoted as $\omega_0$ and the phase shifts are given by $\Delta\phi_{Pv}$ and $\Delta\phi_{Pu}$.
We here briefly summarize the explicit forms of these quantities~\cite{reinken2025unified}.
There are two distinct solutions, which we denote by $+$ and $-$,
\begin{equation}
P_{0\pm} = \sqrt{\frac{6 \nu_\mathrm{v} \tau_\mathrm{d} + 24 \nu_\mathrm{v}- 12 \tau_\mathrm{d} (\gamma_\mathrm{a} \nu_\mathrm{p} - 2 \nu_\mathrm{d}) \pm 2 A}{- 3 \nu_\mathrm{v} \tau_\mathrm{d} + 6 \nu_\mathrm{v}  + 6 \tau_\mathrm{d}  (\gamma_\mathrm{a} \nu_\mathrm{p} + \nu_\mathrm{d}) \mp A}}\, ,
\label{eq:P0pm}
\end{equation}
\begin{equation}
u_{0\pm} = \sqrt{\frac{\nu_\mathrm{p} [3 \nu_\mathrm{v} \tau_\mathrm{d} + 12 \nu_\mathrm{v}  - 6 \tau_\mathrm{d} (\gamma_\mathrm{a} \nu_\mathrm{p} - 2 \nu_\mathrm{d}) \pm A]}{6 \gamma_\mathrm{a} \nu_\mathrm{d}  (\nu_\mathrm{d} \tau_\mathrm{d} + \nu_\mathrm{v})}}\, ,
\end{equation}
\begin{equation}
\omega_{0\pm} = \frac{\sqrt{\nu_\mathrm{d} \tau_\mathrm{d} + \nu_\mathrm{v}}}{\nu_\mathrm{v} \tau_\mathrm{d}}
\sqrt{\frac{12\gamma_\mathrm{a} \nu_\mathrm{p} \nu_\mathrm{d} \tau_\mathrm{d}^2  + [\nu_\mathrm{d} \tau_\mathrm{d} + \nu_\mathrm{v}][3 \nu_\mathrm{v} \tau_\mathrm{d} - 6 \nu_\mathrm{v}  - 6 \tau_\mathrm{d}  (\gamma_\mathrm{a} \nu_\mathrm{p} + \nu_\mathrm{d}) \pm A]}
{-3 \nu_\mathrm{v} \tau_\mathrm{d} + 6 \nu_\mathrm{v}  + 6 \tau_\mathrm{d}  (\gamma_\mathrm{a} \nu_\mathrm{p} + \nu_\mathrm{d}) \mp A}}\, ,
\end{equation}
\begin{equation}
\Delta\phi_{Pu\pm} = \arccos\Bigg(  \sqrt{\frac{-3 \nu_\mathrm{v} \tau_\mathrm{d} + 6 \nu_\mathrm{v}  + 6 \tau_\mathrm{d} (\gamma_\mathrm{a} \nu_\mathrm{p} + \nu_\mathrm{d}) \mp A}{18\gamma_\mathrm{a} \nu_\mathrm{p} \nu_\mathrm{d}  \tau_\mathrm{d}^2/(\nu_\mathrm{d} \tau_\mathrm{d} + \nu_\mathrm{v})}} \Bigg)\, ,
\label{eq:omega0pm}
\end{equation}
\begin{align}
v_{0\pm} &= \sqrt{\frac{\nu_\mathrm{p}^2}{\nu_\mathrm{v}^2}P_{0\pm}^2 + \frac{\nu_\mathrm{d}^2}{\nu_\mathrm{v}^2}u_{0\pm}^2 - 2\frac{\nu_\mathrm{d} \nu_\mathrm{p}}{\nu_\mathrm{v}^2}P_{0\pm} u_{0\pm} \cos(\Delta\phi_{Pu\pm})}\, , \, \\
\Delta\phi_{Pv} &= \arctan\bigg(\frac{\nu_\mathrm{d} u_{0\pm} \sin(\Delta\phi_{Pu\pm})}{\nu_\mathrm{p} P_{0\pm} - \nu_\mathrm{d} u_{0\pm} \cos(\Delta\phi_{Pu\pm})} \bigg)\, ,
\label{eq:v0pm}
\end{align}
where we introduced the abbreviation
\begin{equation}
\begin{aligned}
A = \Big[ &36\gamma_\mathrm{a}^2\nu_\mathrm{p}^2 \tau_\mathrm{d}^2  - 72 \gamma_\mathrm{a} \nu_\mathrm{p} \nu_\mathrm{d} \tau_\mathrm{d}^2  - 36 \gamma_\mathrm{a} \nu_\mathrm{p} \nu_\mathrm{v} \tau_\mathrm{d}^2  - 72 \gamma_\mathrm{a} \nu_\mathrm{p} \nu_\mathrm{v} \tau_\mathrm{d}  + 36 \nu_\mathrm{d}^2 \tau_\mathrm{d}^2   \\
&- 36 \nu_\mathrm{d} \nu_\mathrm{v} \tau_\mathrm{d}^2  + 72 \nu_\mathrm{d} \nu_\mathrm{v} \tau_\mathrm{d}  + 9 \nu_\mathrm{v}^2 \tau_\mathrm{d}^2 - 36 \nu_\mathrm{v}^2 \tau_\mathrm{d}  + 36 \nu_\mathrm{v}^2 \Big]^{\frac{1}{2}}\, .
\end{aligned}
\end{equation}

The red line in Fig.~1 in the main article denotes the boundary of existence of these analytical, spatially uniform, rotational solutions.
When crossing the red line from the disordered, isotropic state by increasing the strength of activity $\nu_\mathrm{p}$, %
the analytical, spatially uniform, rotational solution emerges in a subcritical bifurcation for small enough periodic systems that remain spatially uniform \cite{reinken2025unified}.
Here, the solution denoted by $+$ is stable, whereas the one denoted by $-$ is unstable. Hysteresis emerges.
Decreasing the displacement relaxation time $\tau_\mathrm{d}$, the transition becomes supercritical at a certain point~\cite{reinken2025unified}.
Below that value of $\tau_\mathrm{d}$, the solution denoted by $-$ ceases to exist. Only the solution denoted by $+$ remains.

Moreover, in Fig.~1 of the main article, the upper boundary of the region of hysteresis is indicated by the dashed black line. From here, the disordered isotropic solution becomes linearly unstable, as discussed in the main text. This dashed black line results from the analytical calculations in Sec.~\ref{app:LinearStabilityAnalysis}. %
In contrast to that, the boundary of the hysteretic region for smaller values of $\tau_\mathrm{d}$ is indicated by the dotted black line. It is obtained from numerical simulations, see Sec.~\ref{app:NumericalMethods}.

\section{Numerical methods}
\label{app:NumericalMethods}

We use a pseudo-spectral method~\cite{canuto2007spectral} to solve Eqs.~(1)--(3) in a two-dimensional domain under periodic boundary conditions.
This scheme exploits the fast Fourier transformation to calculate spatial derivatives, which significantly accelerates solving Eq.~(2).
Time integration is performed using a fourth-order Runge--Kutta method and a time step of $\Delta t = 10^{-3}$ in our rescaled units.
To ensure that the velocity field $\mathbf{v}$ stays divergence-free, we employ a projection method~\cite{durran2010numerical}. The pressure $p$ is determined in a way to satisfy incompressibility.

Our numerical calculations start from a weakly perturbed uniform polar state when we investigate the stripe-like patterns shown in Fig.~2(a).
When we address the heterogeneous dynamic states in Fig.~3(a), we instead start from a weakly perturbed isotropic state.
In both cases, these additional random values are taken from a uniform distribution over the interval $[-0.01,0.01]$. They are assigned to each of the components of the polar orientational order parameter field at every grid point in the beginning of the simulation.
Velocity field and memory field of elastic displacements are set to zero initially. 

For the analysis of the heterogeneous states presented in Fig.~3 and the stripe patterns in Fig.~2, we explore large systems of size $200 \times 200$.
In both cases, we use numerical grids of $512 \times 512$ points.
The snapshot of the heterogeneous state in Fig.~3 is obtained for a system of size $100\times 100$.
The data points for the correlation functions shown in Fig.~3(b) represent averages over time intervals of $30$ time units. 
Data points for the correlation length~$\ell$ plotted in the same figure are determined from there and subsequently averaged over $3$ time intervals. 
Error bars in the figure denote standard errors, that is, the calculated standard deviation over all intervals divided by the square root of the number of intervals. Mostly, they are of the size of the data points. 

To obtain the dotted black line in the dynamical state diagram in Fig.~1 of the main article, we perform additional numerical simulations. The region to the left of this line towards the red line is hysteretic. That is, the observed state depends on the initialization of the system. We have crossed the dotted line towards the right, if initialized uniform polar orientational order is not maintained but starts to locally rotate in its fields $\mathbf{P}$, $\mathbf{v}$, and $\mathbf{u}$. %
This we take as a criterion. 
Moving along the dotted line, for corresponding values of the relaxation time $\tau_\mathrm{d}$ we identify the strength of active forcing $\nu_\mathrm{p}$ from where local circling emerges in spite of uniform initialization.
For this purpose, we employ a bisection method, that is, we start from a large interval of possible values of $\nu_\mathrm{p}$ and successively half the size of this interval until an appropriate accuracy is achieved.
Every simulation is initialized with a polar state modulated by the stripe-like pattern presented in the main text.
The system size is always set to twice the most unstable wavelength.
We check for circling states by summing up the angular directional changes of the local polar order.
Although this approach is quite efficient, simulations have to be repeated many times for every value of the relaxation time $\tau_\mathrm{d}$ in order to identify the onset of local circling.
We thus use a lower system size of $64\times64$.

\section{Information on the Supplemental Movies}

The Supplemental Movies show the dynamics in the nonuniform states described in the main article.
To better illustrate how material elements move within the system, the motion of tracers is superimposed.
Here, the black lines follow the evolution of their trajectories.

Supplemental Movie 1 shows the evolution of the vorticity field $\omega = [\nabla \times \mathbf{v}]_z$ resulting from the finite-wavelength instability of the stationary polar state in a periodic system of size $200\times 200$.
The parameters are set to $\gamma_\mathrm{a} = 1$, $\eta = 1$, $\mu = 1$, $\nu_\mathrm{v} = 1$, $\nu_\mathrm{p} = 20$, $\nu_\mathrm{d} = 10$, and $\tau_\mathrm{d} = 0.25$.
Black arrows additionally indicate the velocity field $\mathbf{v}$, demonstrating the nonzero mean velocity and persistent flow due to global polar ordering.
Tracers illustrate the snake-like active motion due to stripe formation.
The duration of the movie covers $4.5$ time units.

Supplemental Movie 2 shows the phase shift $\Delta \phi_{Pv} = \phi_P - \phi_v$, which refers to the local angle between $\mathbf{P}$ and  $\mathbf{v}$ in the nonuniform oscillatory state in a periodic system of size $100\times 100$.
The parameters are set to $\gamma_\mathrm{a} = 1$, $\eta = 1$, $\mu = 1$, $\nu_\mathrm{v} = 1$, $\nu_\mathrm{p} = 20$, $\nu_\mathrm{d} = 10$, and $\tau_\mathrm{d} = 10$.
Again, black arrows mark the velocity field $\mathbf{v}$. These arrows additionally demonstrate the local rotation over time.
The movie is running at half the speed of Supplemental Movie~1 and covers $2.8$ time units.
As there, the motion of a few tracer particles is evaluated, and the black lines follow the evolution of their trajectories.
They do not describe perfect circular loops as in a uniform rotating state. Instead, their trajectories depend on the location within  the evolving domains.
For solid-like materials ($\tau_\mathrm{d} \to \infty$), the areas explored by the tracers are limited in size, due to the role of the field of memorized displacements $\mathbf{u}$ and the resulting elastic restoring forces.
In the movie, the relaxation time is set to an elevated value of $\tau_\mathrm{d} = 10$.

Supplemental Movie 3 shows the dynamics of topological defects of the velocity field $\mathbf{v}$ (not shown explicitly), superimposed on the phase shift $\Delta \phi_{Pv} = \phi_P - \phi_v$ (color code), in a periodic system of size $100\times 100$.
Positive topological $+1$ defects, which are located in the center of vortices of the velocity field, are shown as red dots, whereas negative topological $-1$ defects, located in between, are shown as blue dots.
The movie visualizes the pairwise creation and annihilation of defects, as well as their dynamic motion.
As in Supplemental Movie 2, the parameters are set to $\gamma_\mathrm{a} = 1$, $\eta = 1$, $\mu = 1$, $\nu_\mathrm{v} = 1$, $\nu_\mathrm{p} = 20$, $\nu_\mathrm{d} = 10$, and $\tau_\mathrm{d} = 10$.
The movie is running at half the speed of Supplemental Movie~2 and covers $0.75$ time units.

\section{Topological defects}

\begin{figure*}
\includegraphics[width=0.999\linewidth]{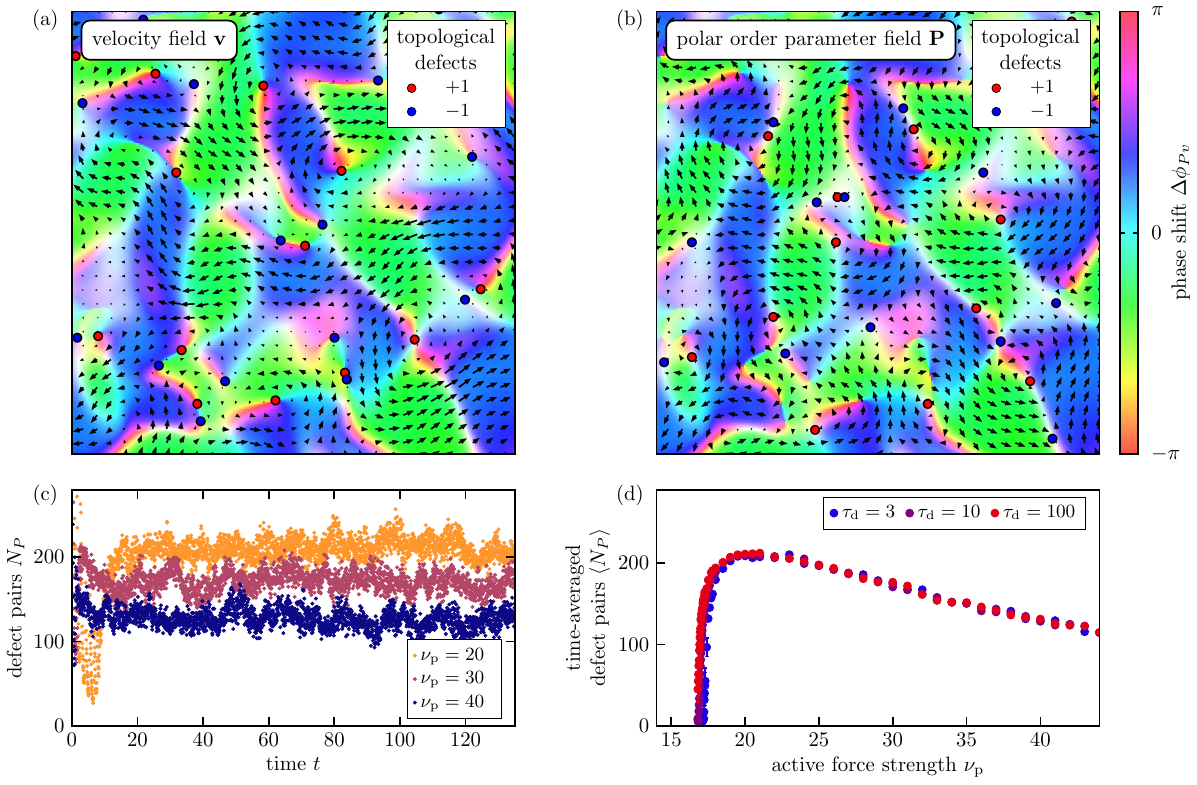}
\caption{\label{fig:Defects}Topological defects in (a) the velocity field $\mathbf{v}$ (black arrows) and (b) the polar order parameter field $\mathbf{P}$ (black arrows), superimposed on the field of phase shift $\Delta \phi_{Pv} = \phi_P - \phi_v$ (color code).
Positive $+1$ defects are located in the center of vortices of each field and marked as red dots.
Negative $-1$ defects are located in between and indicated by blue dots.
The snapshots show a $50 \times 50 $ window of a larger periodic system of size $100 \times 100$ and results from simulations for an active forcing of strength $\nu_\mathrm{p} = 20$ and a viscoelastic relaxation time $\tau_\mathrm{d} = 10$.
(c) Time evolution of the number of defect pairs $N_P$ in the polar orientational order parameter field $\mathbf{P}$ when starting from initial conditions of disordered states for different values of the strength of active forcing $\nu_\mathrm{p}$ and a viscoelastic relaxation time $\tau_\mathrm{d} = 10$ in a system of size $200\times 200$.
(d) Time-averaged number of defect pairs $\langle N_P\rangle$ in the polar order parameter field $\mathbf{P}$ as a function of $\nu_\mathrm{p}$ for different values of $\tau_\mathrm{d}$.
Remaining parameter values are set to  $\gamma_\mathrm{a} = 1$, $\eta = 1$, $\mu = 1$, $\nu_\mathrm{v} = 1$, and  $\nu_\mathrm{d} = 10$. }
\end{figure*}

Topological defects are singularities, for example, in the fields of orientational order in nematic or polar phases.
These defects are characterized by their topological ``charges''. They can be obtained by counting the number of rotations of the direction of local orientational order when circumnavigating the defect.
For example, in the case of polar orientational order, defects of ``charge'' $+1$ describe the center of a vortex, whereas defects of ``charge'' $-1$ are located in between vortices.
The concept of topological defects is useful in the context of active matter, because aspects of collective behavior of the self-propelled units are quantified by their degree of orientational ordering.
In the case of active nematics, defects show half-integer ``charges'' due to the nematic symmetry of the local orientational order. That is, we recover the local nematic axis already by half of a full rotation.
The dynamics of defects in nematic orientational order is intimately connected to the emergence of active nematic turbulence~\cite{giomi2013defect,doostmohammadi2018active,alert2021active}.
In systems characterized by polar order~\cite{rana2024defect}, topological ``charges'' show integer values. Only after a full rotation we recover the local direction of polar orientational order.

Our study of systems of polar orientational order falls into the latter class.
Thus, we only encounter integer defects.
Besides defects in the polar orientational order parameter field $\mathbf{P}$, we consider defects in the velocity field $\mathbf{v}$, which is polar as well.
In order to identify the defects, we determine the local angles of orientation of $\mathbf{v}$ and $\mathbf{P}$.
A point in space is then identified as a defect if the increments of the angular orientation of the field, when circling around the point, sum up to a nonzero value.
This value gives the topological ``charge'' of the defect.
Figures~\ref{fig:Defects}(a) and (b) show snapshots of the velocity and associated polar order parameter field, with defects marked by red ($+1$) and blue ($-1$) dots.
The background color indicates the phase shift between the fields, $\Delta \phi_{Pv} = \phi_P - \phi_v$, as in Fig.~3 in the main article.
Since the two fields $\mathbf{v}$ and $\mathbf{P}$ are not necessarily parallel but show distinct patterns of vortices, the locations of the defects in (a) and (b) are not identical.

In order to characterize the emerging spatial structures, we determine the number of defect pairs within the system.
Figure~\ref{fig:Defects}(c) shows the time evolution of the number of defect pairs $N_P$ of the polar order parameter field for different values of the strength of active forcing $\nu_\mathrm{p}$ at $\gamma_\mathrm{a} = 1$, $\eta = 1$, $\mu = 1$, $\nu_\mathrm{v} = 1$,  $\nu_\mathrm{d} = 10$, and $\tau_\mathrm{d} = 10$.
Here, we started the simulations from an initially isotropic state with random local perturbations, see Sec.~\ref{app:NumericalMethods}.
After an initial transient, $N_P$ fluctuates around a constant value depending on $\nu_\mathrm{p}$.
This constant value is obtained as a time average $\langle N_P \rangle$ and plotted in Fig.~\ref{fig:Defects}(d) as a function of $\nu_\mathrm{p}$.
As for the time-averaged number of defects in the velocity field $\langle N_v\rangle$ discussed in the main text and there shown in Fig.~3(e), $\langle N_P \rangle$ likewise behaves oppositely to the domain size $\ell$, see Fig.~3(c) in the main article.
Thus, larger domain sizes $\ell$ correspond to lower numbers in topological defect pairs $\langle N_v \rangle$ and $\langle N_P \rangle$, and vice versa.

\section{Rheologically tuned transport of embedded cargo}

We perform additional numerical simulations to obtain trajectories of passive cargo objects under the influence of varying rheological properties, see Fig.~4.
To this end, we periodically change the relaxation time $\tau_\mathrm{d}$ of the elastic memory field of the active carrier medium. Thus, we alternately increase and decrease the influence of elasticity.
In detail, we switch between $\tau_\mathrm{d}^\mathrm{l}$ and $\tau_\mathrm{d}^\mathrm{h}$ using a smooth periodic function  
\begin{equation}
\tau_\mathrm{d}(t) = \frac{\tau_\mathrm{d}^\mathrm{l}}{2} + \frac{\tau_\mathrm{d}^\mathrm{h}}{2} - \frac{(\tau_\mathrm{d}^\mathrm{h} - \tau_\mathrm{d}^\mathrm{l})}{\pi} \arctan\bigg(\frac{\sin(2 \pi t f_\mathrm{ex} + \pi/3)}{\delta}\bigg)\, .
\end{equation}
Here, $f_\mathrm{ex}$ is the oscillation frequency and $\delta$ determines how quickly the transitions take place. 
The results presented in the main article are obtained for $\tau_\mathrm{d}^\mathrm{l} = 0.25$, $\tau_\mathrm{d}^\mathrm{h} = 1$, $f_\mathrm{ex} = 0.2$, and $\delta = 0.05$.
For the other parameter values, we set $\gamma_\mathrm{a} = 1$, $\eta = 1$, $\mu = 1$, $\nu_\mathrm{v} = 1$, $\nu_\mathrm{d} = 10$, and $\nu_\mathrm{p} = 20$.
Furthermore, we fix the system size to $100\times 100$ and use a spatial resolution of $256\times 256$.
The passive cargo units are considered to be transported under low-Reynolds-number conditions and thus follow the velocity field $\mathbf{v}$. %

\newpage


%